\documentclass[prl,aps,twocolumn,showpacs,preprintnumbers,superscriptaddress,amsmath,amssymb]{revtex4-1}
\usepackage{graphicx}
\usepackage{dcolumn}
\usepackage{bm}
\usepackage{hyperref}

\begin{document}

\title{Metal-insulator Transition in VO$_{2}$: a DFT+DMFT perspective}

\author{W. H. Brito}
\address{Departamento de  F\'{\i}sica, Universidade  Federal de Minas Gerais, C. P. 702, 30123-970, Belo Horizonte, MG, Brazil.}
\address{Department of Physics and Astronomy, Rutgers University, Piscataway, New Jersey 08854, USA.}
\author{M. C. O. Aguiar}
\address{Departamento de  F\'{\i}sica, Universidade  Federal de Minas Gerais, C. P. 702, 30123-970, Belo Horizonte, MG, Brazil.}
\author{K. Haule}
\author{G. Kotliar}
\address{Department of Physics and Astronomy, Rutgers University, Piscataway, New Jersey 08854, USA.}


\begin{abstract} 

We present a theoretical investigation of the electronic structure of rutile (metallic) and M$_1$ and M$_2$ monoclinic (insulating) phases of VO$_2$ employing a fully self-consistent combination of density functional theory and embedded dynamical mean field theory calculations. We describe the electronic structure of the metallic and both insulating phases of VO$_2$, and propose a distinct mechanism for the gap opening.
We show that  Mott physics plays an  essential role in all phases of VO$_2$: undimerized vanadium atoms undergo classical Mott transition through local moment formation (in the  M$_2$ phase), while strong superexchange within V-dimers adds significant dynamic intersite correlations, which remove the singularity of self-energy for dimerized V-atoms. The resulting transition from rutile to dimerized M$_1$ phase is adiabatically connected to Peierls-like transition, but is better characterized as the Mott transition in the presence of strong intersite exchange. As a consequence of Mott physics, the gap in the dimerized M$_1$ phase is temperature dependent. The sole increase of electronic temperature collapses the gap, reminiscent of recent experiments.

\end{abstract}


\maketitle

VO$_2$ is  of great   practical importance as its MIT is very close to room temperature, and its ultrafast switching between metallic and insulating phases can be used for device applications~\cite{ramanathanAnnuRev}.  VO$_2$ is also of fundamental interest, due to the complex interplay between electronic correlations and structural distortions, and as a result the physical mechanism responsible for the gap formation has remained under debate.

Structurally, VO$_2$ undergoes a transition from a high temperature rutile phase~\cite{mcwhan} (R) to low temperature monoclinic phases. In the latter, under ambient pressure, the vanadium atoms dimerize and tilt with respect to the rutile $c$ axis, giving rise to the M$_1$ phase~\cite{jmlong}. Due to the dimerization it is expected that the electronic states associated with the overlapping $d$-orbitals along the rutile $c$ axis ($a_{1g}$ states) split into bonding-antibonding subbands. In addition, the antiferroelectric displacement leads to an upshift of the subbands associated with the remaining $t_{2g}$ states ($e_{g}^{\pi}$). As a result, a gap would appear between the $a_{1g}$ and $e_{g}^{\pi}$ subbands, suggesting thus that VO$_2$ undergoes a Peierls-type transition~\cite{goodenough}. However, density functional theory (DFT) calculations showed that these structural distortions alone cannot give rise to a gap in the M$_1$ phase~\cite{wentzprl,eyertAnnPhys}. In addition, a distinct monoclinic phase (M$_
2$) appears when the system is under uniaxial stress or doped with Cr$^{3+}$, Al$^{3+}$, Fe$^{3+}$, or Ga$^{3+}$~\cite{pouget2,strelcov}. In this phase half of the vanadium atoms pair along the rutile $c$ axis, while the other half experience a zigzag-like distortion along the same axis. In particular, the existence of localized electrons in the zigzag chains also suggests that electronic correlations indeed play a role in the gap formation of VO$_2$.

The cluster-dynamical mean field theory (DMFT) has successfully described several aspects of the physics of VO$_2$, but some discrepancies between different implementations  remain. Biermann \textit{et al.}~\cite{biermannPRL} showed that the M$_1$ phase is insulating, with the gap of 0.6 eV in agreement with experimental findings~\cite{koethePRL}. Furthermore, the authors found that electronic intersite correlations, within the vanadium dimers, renormalize down the $a_{1g}$ bonding-antibonding splitting in comparison with DFT calculations. In their proposed mechanism the M$_1$ phase can be viewed as a renormalized Peierls insulator. In contrast, by means of \textit{ab initio} linear scaling DFT+cluster-DMFT calculations, Weber \textit{et al.}~\cite{cedricPRL} observed that the gap formation of the M$_1$ phase is mainly due to an orbital-selective Mott instability concerning the $a_{1g}$ electronic states in our notation. Looking at the occupancy of the $3d$ shell, they obtained at around two electrons per 
vanadium, resulting in four electrons per vanadium dimer, suggesting that the M$_1$ phase is \textit{not} a renormalized Peierls insulator~\cite{biermannPRL}. 

Important fundamental challenges remain: i) the close proximity of the M$_1$, M$_2$, and R phase in the phase diagram~\cite{strelcov,park} calls for an unified framework in which these phases can be simultaneously described; ii) another fundamental question is whether the MIT  transitions will take place  if we fix the structure and change only the temperature.

It has not been possible to properly address these issues within an approach based on an effective model from which the oxygen degrees of freedom are eliminated, and which contains only V-$3d$ electrons (Hubbard model).  
Indeed, earlier DMFT works on the Hubbard model~\cite{biermannPRL,jmtomczak,lazarovits} showed that for a given set of Hubbard $U$ and $J$ parameters, the gap in the M$_1$ insulating phase is too robust, persisting to very high temperatures~\cite{biermannPRL}, while the mass renormalization in the R phase is too small compared to experiment~\cite{lazarovits,jmtomczak}. Both of these effects result from placing this material too far from the Mott transition boundary. Consequently, the M$_2$ phase was not described before with this approach, as half of V-atoms would not undergo Mott-Hubbard transition.

In this letter we solved these problems by describing VO$_2$ with modern all electron embedded DMFT approach, where in addition to correlated vanadium atoms the itinerant states of oxygen are included in the Dyson self-consistent equation~\cite{hauleWK,footnote1}.  We present a comprehensive picture and describe all the phases of VO$_2$ with the same values of the  ($U$,$J$) parameters, including the M$_2$ phase which has not been considered previously in DMFT treatments. Mott physics is central for the proper description of all the phases even though the Mott instability is arrested in the M$_1$ phase.

Our theory leads to a different physical picture for the gap opening in monoclinic phases of VO$_2$, and most importantly the possibility of the \textit{collapse} of the M$_1$ insulating state by temperature. Our results indicate the presence of significant intersite correlations within the vanadium dimers, which in turn lower the $a_{1g}$ subband in relation to the $e_{g}^{\pi}$. In particular, we notice that nonlocal dynamic correlations \textit{enhance} the $a_{1g}$ bonding-antibonding splitting, in contrast to what was reported by Biermann~\textit{et al.}~\cite{biermannPRL}.  
The electrons in all phases of VO$_2$ are in the near vicinity of the Mott transition, but the true pole in the self-energy, signalizing the local moment formation, occurs only in the paramagnetic M$_2$ phase on undimerized V-atoms and its $a_{1g}$ orbital. In the M$_1$ phase and in the antiferro-ordered M$_2$ phase, the singularity of the self-energy is arrested as the ordered states are adiabatically connected to Peierls and Slater insulators, respectively. 
The adiabatic connection between the weakly and the corresponding strongly interacting states, makes the Mott mechanism hard to distinguish from alternative scenarios, nevertheless, collapsing a large insulating gap with raising electronic temperature is possible only in the Mott state in the presence of strong superexchange, as was found for example in the cluster-DMFT study of the 2D Hubbard model~\cite{hpark}. Hence, according to our results the M$_1$ phase is best characterized as the Mott phase in the presence of strong intersite superexchange within the V-dimers, while the undimerized V-atoms in the M$_2$ phase undergo canonical Mott transition associated with local moment formation in the presence of weak superexchange.

\begin{figure}[!htb]
\includegraphics[scale=0.37]{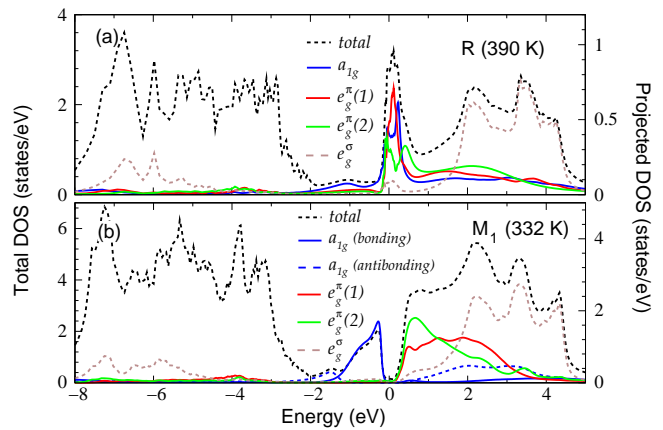}
 \caption{(Color online) DFT+DMFT-based total (black dashed line) and projected DOS of (a) R and (b) M$_{1}$ phase of VO$_2$. The projections to $a_{1g}$, $e_{g}^{\pi}(1)$, $e_{g}^{\pi}(2)$, and $e_{g}^{\sigma}$ states are shown in blue, red, green, and brown lines, respectively. For the M$_1$ phase, the solid (dashed) blue line corresponds to the projection on the bonding (antibonding) $a_{1g}$ molecular state.}
\label{fig:dos_RaM1}
\end{figure}

\textit{Rutile and M$_1$ phase.--} We first address the R phase at temperature of 390 K. In Fig.~\ref{fig:dos_RaM1}(a) we show the calculated total, $t_{2g}$ and $e_{g}^{\sigma}$ projected density of states (DOS).

The R phase is metallic with strongly renormalized $t_{2g}$ orbitals compared to corresponding DFT results (not shown).  A weak lower Hubbard band (LHB) is observed at around -1.09 eV, and a broad upper Hubbard band (UHB) around 2.54 eV. The former is in good agreement with experimental photoemission measurements~\cite{koethePRL}, and previous theoretical findings~\cite{biermannPRL}. The LHB and UHB are mainly of $a_{1g}$ and $e_{g}^{\pi}$ character, respectively.  

We next present the electronic structure of the M$_1$ phase. Within the V-dimer, treated as cluster in DMFT, it is useful to adopt the symmetric and antisymmetric combination of orbitals. The associated self-energies are denoted as the bonding $\Sigma_{b,\alpha}$ and antibonding $\Sigma_{ab,\alpha}$ components, where $\alpha = \{a_{1g},e_{g}^{\pi}(1),e_{g}^{\pi}(2)\}$~\cite{footnote1}.
In Fig.~\ref{fig:dos_RaM1}(b) the total, $t_{2g}$, and $e_{g}^{\sigma}$ projected DOS are shown. At first, we notice the opening of a gap of 0.55 eV, between the $a_{1g}$ and $e_{g}^{\pi}$ subband, which is in good agreement with experimental and previous theoretical findings~\cite{koethePRL,biermannPRL,cedricPRL}. As expected, upon dimerization and antiferroelectric distortion, the $a_{1g}$ subband splits in bonding (solid blue) and antibonding (dashed blue) states, while the $e_{g}^{\pi}$ subband upshifts in comparison with the R phase. The $a_{1g}$ bonding orbital has a coherent peak at around -0.30 eV, while the antibonding orbital has two incoherent peaks reminiscent of LHB and UHB located at -1.5 eV and 2.58 eV, respectively. The coherent peak and the satellite below it were also observed in previous theoretical~\cite{biermannPRL,lazarovits} and experimental~\cite{koethePRL} works. However, in our case, the UHB does not represent a weak coherent peak associated with the antibonding state, as in Ref.~\
cite{biermannPRL}. 
The  bonding-antibonding splitting (relative to the DFT and DFT + single-site DMFT calculations (not shown)) increases upon inclusion of intersite dynamic correlations within the dimer, in contrast to Ref.~\cite{biermannPRL}.  

\begin{figure}[!htb]
 \includegraphics[scale=0.75]{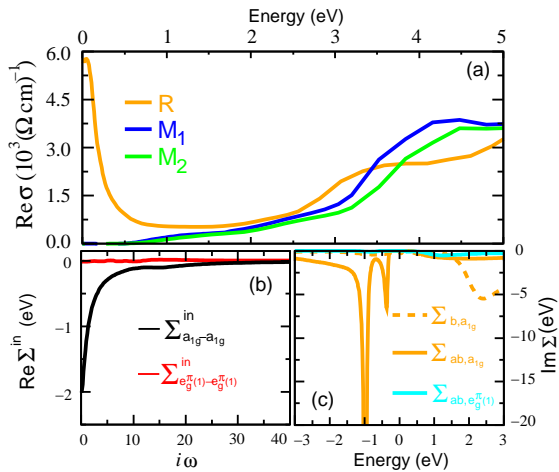}
\caption{(Color online) (a) Real part of the optical conductivity of R, M$_1$, and M$_2$ phases of VO$_2$. (b) Real part of intersite self-energies, on imaginary frequency axis, of $a_{1g}$-$a_{1g}$ (black) and $e_{g}^{\pi}(1)$-$e_{g}^{\pi}(1)$ (red) states of M$_1$ phase. (c) Imaginary part, on real frequency axis, of bonding (dashed lines) and antibonding (solid lines) self-energies associated with the $a_{1g}$ (orange) and $e_{g}^{\pi}(1)$ (cyan) dimer electronic states.}
\label{fig:self_optic}
\end{figure}

We also computed the optical conductivity of R, M$_1$, and M$_2$ phases of VO$_2$, which are presented in Fig.~\ref{fig:self_optic}(a).
We calculate the plasma frequency of the R phase, defined as $\omega_p^2 ={2}/{(\pi\varepsilon_0)}\int_0^{\Lambda} \sigma(\omega)d\omega$, with $\Lambda=2\,$eV and $\Lambda=0.1\,$eV, the latter denoted by ${\omega_p^*}^2$. At $390$~K we obtained for the ratio $\omega_p^2(LDA)/\omega_p^2(DMFT)=2.24$, and $\omega_p(DMFT)=3.28\,$eV, in good agreement with Ref.~\cite{okazaki}.                                             
We notice a mild temperature dependence of this ratio, decreasing with decreasing temperature. This signals an increase of kinetic energy due to gain of coherence, characteristic of Mott-Hubbard system. On the other hand, the ratio ${\omega_p^*}^2(LDA)/{\omega_p^*}^2(DMFT)$ is much larger: at $100$~K it is 4.71, and increases as the Drude peak broadens (at $390$~K it is 6.12). Finally, the DMFT mass enhancements, extracted from the slope of the self-energy, are $m^*/m_{band}=3.5,3,2.5$ for $a_{1g}$, $e_g^{\pi}(1)$, and $e_g^{\pi}(2)$, respectively. In the conventional Hubbard model picture, the mass enhancement ($m^*/m_{band}$) and the ratios of plasma frequencies ($\omega_p^2(LDA)/\omega_p^2(DMFT)$) should be very similar, and both are used to measure the correlation strength in the literature~\cite{qazilbash1}. In VO$_2$ these measures of correlations differ because the plasma frequency $\omega_p$ includes substantial contribution from the interband transitions, which strongly mix with less correlated 
oxygen and $e_g$ states, resulting in apparently weaker correlations than given by $m^*_{t2g}/m_{band}$, a physics beyond Hubbard model. The low energy ratio of ${\omega_p^*}^2$ has negligible interband contribution, and at zero temperature it should be equal to  $m^*_{t2g}/m_{band}$, but at finite temperature it gets large due to broadening and the loss of strength of the Drude peak.

\textit{Effects of nonlocal dynamic correlations.--} Having described the electronic structure of both R and M$_1$ phases of VO$_2$, we next investigate how nonlocal dynamic correlations contribute to the opening of a gap between $a_{1g}$ and $e_{g}^{\pi}$ subbands in the M$_1$ phase. Fig.~\ref{fig:self_optic}(c) shows the bonding and antibonding components of the imaginary part of the self-energy associated with the $a_{1g}$ and $e_{g}^{\pi}(1)$ dimer electronic states.

We notice that once the Peierls instability occurs in our calculation, the Mott instability is arrested, hence there is no pole in the imaginary part of the self-energy associated with the $a_{1g}$ or $e_{g}^{\pi}$ states. This excludes an orbital-selective Mott-Hubbard mechanism as the source of the gap of the M$_1$ phase, as proposed in Ref.~\cite{cedricPRL}. Our results bear strong resemblance with the Mott transition of the Hubbard model in cluster-DMFT~\cite{haule_cluster}, where the local singlet state of the cluster dominates the low energy properties of the model. The energy gain to form the strong bonding state on the cluster is here not just due to increased hopping between the two V-atoms, but it is primarily due to the gain of the exchange energy, which is stronger than kinetic energy, as the latter is strongly reduced due to proximity to the Mott transition. The precise calculation of this exchange energy is beyond this work since it requires the computation of the momentum dependent spin-susceptibility which in turn depends on the two particle vertex function.

Next we define the local and intersite self-energies, which are related to the self-energies in the dimer basis as $\Sigma^{local} = \frac{1}{2}(\Sigma_{b}+\Sigma_{ab})$ and $\Sigma^{in(tersite)} = \frac{1}{2}(\Sigma_{b}-\Sigma_{ab})$~\cite{footnote1}. In Fig.~\ref{fig:self_optic}(b) we also show the real part of the intersite components of self-energies associated with the same electronic states as those in the inset. We observe that the component of $e_{g}^{\pi}(1)$ is negligible, but, notably, one can see that $Re \Sigma_{a_{1g}-a_{1g}}^{in}$ depends strongly on the frequency in the low-energy part. This indicates the presence of strong intersite electronic correlations within the vanadium dimers, which in turn lower the $a_{1g}$ bonding state. As a result, the bonding-antibonding splitting \textit{increases} and a gap between the $a_{1g}$ and $e_{g}^{\pi}$ appears (for more details about the bonding-antibonding splitting see the supplemental material). 
Thus, one can consider that nonlocal correlations give rise to an effective $a_{1g}-a_{1g}$ frequency dependent hopping $t_{a_{1g}-a_{1g}} + Re\Sigma_{a_{1g}-a_{1g}}^{in} (i\omega)$, which properly takes into account the $a_{1g}$ bonding-antibonding splitting.  A similar observation was previously proposed for the low-temperature phase of Ti$_2$O$_3$~\cite{poteryaev}, but required a strong intersite Coulomb interaction for opening the gap.

\textit{Metallization due to hot carriers.--} More recently, numerous experimental studies have reported the existence of monoclinic-like metallic phases of VO$_2$~\cite{kim, arcangeletti,tao,laverockMM,wegkamp}. 
This transition was also induced by the application of femtosecond laser pulses on VO$_2$ films. As pointed out by Wegkamp~\textit{et al.}~\cite{wegkamp}, the photoexcitation gives rise to hot carriers, which have an associated temperature much higher than the lattice temperature. 
Motivated by this fact, we performed calculations considering a much higher temperature, \textit{i.e.} $T = 900$~K, for electrons in the M$_1$ phase. 

\begin{figure}
  \includegraphics[scale=0.6]{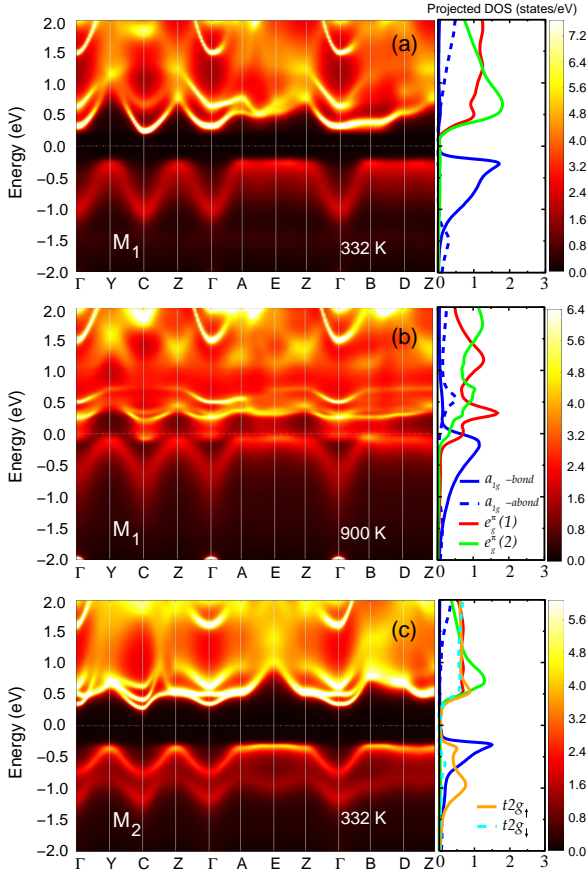}
  \caption{(Color online) Spectral function and projected DOS of M$_1$ phase at $332$ (a) and $900$ K (b). In (c) the spectral function of M$_2$ phase at $332$ K is shown.}
\label{fig:akw_Ht}
\vspace{-10pt}
\end{figure}

\begin{figure}[!htb]
 \includegraphics[scale=0.6]{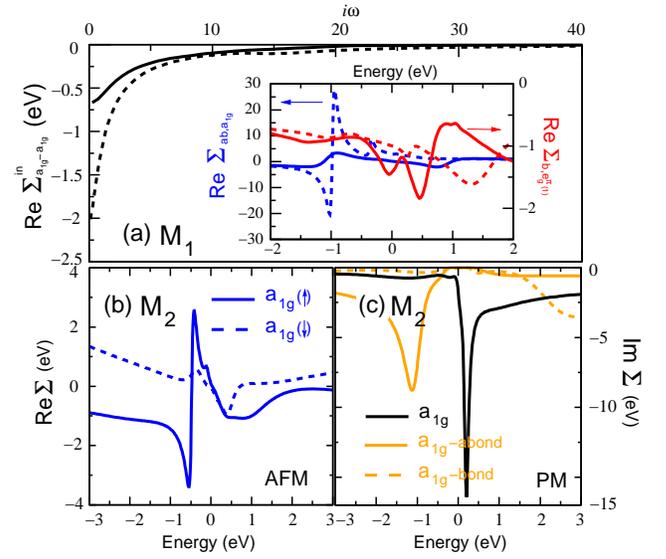}
\caption{(Color online)(a) $Re\Sigma_{a_{1g}-a_{1g}}^{in}(i\omega)$ at $332$~K (dashed lines) and $900$ K (solid lines) of M$_1$ phase. Inset: $Re \Sigma_{ab,a_{1g}}(\omega)$ (blue) and $Re\Sigma_{b,e_{g}^{\pi}(1)}(\omega)$ (red) at $332$ (dashed lines) and $900$ K (solid lines). (b) $Re \Sigma_{a_{1g}}(\omega)$ of undimerized V-atoms in the antiferromagnetic phase of M$_2$ phase. (c) $Im \Sigma(\omega)$ of bonding and antibonding $a_{1g}$ self-energies of V-dimers (orange lines) and $a_{1g}$ states of undimerized V-atoms (black lines).}
\label{fig:self_m1m2}
\end{figure}

In Fig.~\ref{fig:akw_Ht}(a) and (b) we show the calculated spectral function and the projected DOS of M$_1$ at $332$ and $900$~K, respectively. 
From the spectral function at $900$~K, one can see the closing of the gap. 
In particular, we observe that $a_{1g}$ and $e_{g}^{\pi}(1)$ subbands shift towards the Fermi level, the latter shifting more than the $a_{1g}$ subband. 
To understand why the gap vanishes at $900$ K, we examine the temperature dependence of the self-energies. In Fig.~\ref{fig:self_m1m2}(a) we present the real part of intersite $\Sigma^{in}_{a_{1g}-a_{1g}}$ for these two temperatures. In addition, in the inset we show the antibonding component of $Re \Sigma_{ab,a_{1g}}$ and the bonding component of $Re \Sigma_{b,e_{g}^{\pi}(1)}$, on the real frequency axis. Here, we observe that $Re \Sigma_{a_{1g}-a_{1g}}^{in}$, in the low-energy part, and $Re \Sigma_{ab,a_{1g}}$ are strongly suppressed with increasing temperature. Therefore at $900$ K the renormalization of the $a_{1g}$ subband decreases significantly in comparison to that at $332$ K. Thus, this subband is shifted towards the Fermi level. In relation to the $e_{g}^{\pi}(1)$ subband, we note an enhancement of the $Re \Sigma_{b,e_{g}^{\pi}(1)}$ from -0.46 eV to 0.77 eV, leading to an downshift of this subband.

\textit{M$_2$ phase.--} Our cluster-DMFT simulation of the M$_2$ phase, using the same values of $U$ and $J$ employed for the other phases, are presented in Fig.~\ref{fig:akw_Ht}(c). The size of the insulating gap is $\approx$ 0.58 eV. In addition we show in Fig.~\ref{fig:self_optic}(a) the calculated optical conductivity of M$_2$ phase. Interestingly, we observe that both monoclinic phases are very similar, providing further support for the Mott point of view.

For this simulation we allowed anti-ferromagnetic ordering of undimerized vanadium atoms, but we find a similar gap also in paramagnetic simulation of this phase.
In both simulations, we find that the $a_{1g}$ orbital on the dimerized V-atoms is renormalized due to intersite correlations, in similar way as in the M$_1$ phase. However, the splitting mechanism of the $a_{1g}$ orbital on undimerized V-atoms depends on the magnetic ordering. In particular, in the anti-ferromagnetic phase the $a_{1g}$ orbital is renormalized by the spin-dependent real part of the local $a_{1g}$ self-energy, which depends strongly on  frequency in the low-energy part as shown in Fig.~\ref{fig:self_m1m2}(b). In contrast, in the paramagnetic phase the $a_{1g}$ self-energy has a pole in the imaginary part as can be seen in Fig.~\ref{fig:self_m1m2}(c). This indicates that in the paramagnetic phase, the $a_{1g}$ orbital of the undimerized V-atoms splits due to the canonical Mott instability, proving that the M$_2$ phase should be characterized as a Mott insulator. In addition, this Mott instability also leads to an orbital polarization in favor of the $a_{1g}$ orbital of undimerized V atoms by about 0.34/V atom with respect to the same orbital in R phase. It is noteworthy that the $e_{g}^{\pi}$ orbital occupation decreases almost by the same amount (-0.28/V atom).  Finally, we mention that this orbital polarization leads to the stabilization of the $a_{1g}$ electronic states and therefore cooperates for the stabilization of the insulating state in the M$_2$ phase.

\textit{Conclusions.--} 
Our work highlights the importance of describing simultaneously all phases of a strongly correlated material.
The importance of Mott physics in all the phases of VO$_2$ was stressed early on in the pioneering work of Pouget and Rice~\cite{RFS}. This physics now emerges from a quantitative first principles method, and its implications for many physical quantities has been elucidated.

\begin{acknowledgments}
We acknowledge support from the Brazilian agencies CNPq, FAPEMIG and CAPES. W.H.B. and M. C. O. A. acknowledge A. M. de Paula and V. B. Nascimento for useful discussions. K.H. and G.K. were supported by NSF DMR-1405303 and NSF DMR-1308141, respectively.
\end{acknowledgments}

\end{document}


\title{Supplemental Material for \\ ``Metal-insulator Transition in VO$_{2}$: a DFT+DMFT perspective''}

\author{W. H. Brito}
\address{Departamento de  F\'{\i}sica, Universidade  Federal de Minas Gerais, C. P. 702, 30123-970, Belo Horizonte, MG, Brazil.}
\address{Department of Physics and Astronomy, Rutgers University, Piscataway, New Jersey 08854, USA.}
\author{M. C. O. Aguiar}
\address{Departamento de  F\'{\i}sica, Universidade  Federal de Minas Gerais, C. P. 702, 30123-970, Belo Horizonte, MG, Brazil.}
\author{K. Haule}
\author{G. Kotliar}
\address{Department of Physics and Astronomy, Rutgers University, Piscataway, New Jersey 08854, USA.}

\maketitle

\section{Method}

We performed all-electron charge self-consistent DFT+DMFT
calculations, with implementation described in
Ref.~\cite{hauleWK}. This implementation avoids downfolding, and
rather extremises DMFT functional in real space.  This implementation
has been tested in numerous materials, such as transition metal
oxides~\cite{deng,turan,zhang_iridates,ohta, stewart}, iron-based
superconductors~\cite{zpyin,mandal,lee_ir},
actinides~\cite{janoschek,zhu,deng_pu},
lanthanides~\cite{chakrabarti,choi1,choi2}, and transition metal
compounds~\cite{mazumdar,mandal2}. Here it is applied, for the first
time, to a system that requires cluster-DMFT treatment.

In our calculations we included dynamic correlations within
single-site and also within two-site cluster on dimerized V-atoms (in
monoclinic phases). In particular, the single-site DMFT is very
accurate for non-dimerized vanadium atoms (see Refs.
~\cite{biermannPRL,lazarovits}), while the dimerized sites require
cluster-DMFT treatment.  Within our calculations, we obtain the
electronic charge density, chemical potential, impurity levels,
self-energy ($\Sigma$), and Green's function self-consistently.  We
mention that the full charge self-consistency considerably improves on
one-shot calculation, as has been for example shown in the case of
metal-insulator transition in V$_2$O$_3$~\cite{leonov_v2o3}. The
all-electron DFT calculations were performed within
Perdew-Burke-Ernzerhof generalized gradient approximation
(PBE-GGA)~\cite{pbe}, as implemented in Wien2K code~\cite{wien}. The
quantum impurity problem was solved by means of Continuous time
quantum Monte Carlo (CTQMC) calculations~\cite{ctqmc}, at fixed
distinct temperatures for R and monoclinic phases. To compute the
self-energy on the real frequency axis we performed analytic
continuation using the maximum entropy method. Throughout our calculations we employed the
Coulomb interaction $U = 6.0$ eV and Hund's coupling $J = 1.0$ eV. We
have also used the local axis system proposed by
Eyert~\cite{eyertAnnPhys}, wherein the $d_{x^{2}-y^{2}}$ orbitals
mediate the $\sigma$-type $d-d$ overlap along the rutile c axis, while
the remaining $t_{2g}$ orbitals, \textit{i. e.} $d_{xz}$ and $d_{yz}$,
mediate the $\pi$-type $d-d$ overlap. In our notation, the former is
denoted as $a_{1g}$ and the latter as the doubly degenerated
$e_{g}^{\pi}$ states. In particular, in our investigation
$\{e_{g}^{\pi}(1),e_{g}^{\pi}(2)\}$ denotes the $\{d_{xz},d_{yz}\}$
and $\{d_{yz},d_{xz}\}$ orbitals in R and M$_1$ phase,
respectively. In our calculations only the $t_{2g}$ states were
considered in the impurity model, while the remaining $e_{g}^{\sigma}$
states were treated within DFT (GGA). Finally, to investigate the R
and M$_1$ phase we used the experimental structural parameters from
Ref.~\cite{mcwhan} and Ref.~\cite{jmlong}, respectively.

\section{Dyson equation of the solid}

The correlations are quite local in real space, but they are not very
local in a Wannier representation, which eliminated oxygen degrees of
freedom and builds Hubbard model, keeping vanadium d-bands only. The
DMFT (c-DMFT) approximation, which treats exactly the dynamic
correlations on a give atom (or cluster), is not a very good
approximation in the latter case, because Wannier functions are quite
delocalized and extending on oxygen sites. On the other hand, DMFT is
an excellent approximation in real space, when dynamic correlations are
expanded in terms of quasi-localized atomic orbitals, i.e.,
\begin{eqnarray}
\Sigma_{i\omega}(\vr,\vr') = \sum_{mm',\mu\mu'} \braket{\vr|\Phi^\mu_m}
\braket{\Phi^\mu_m | \Sigma |\Phi^{\mu'}_{m'}}
\braket{\Phi^{\mu'}_{m'}|\vr'}
\label{Eq:Sigma}
\end{eqnarray}
where $\ket{\Phi^\mu_m}$ form a complete basis set around vanadium atoms
centered at $\mu$, and $m,m'$ enumerate atomic degrees of freedom. Our
quasi-atomic orbitals vanish outside the radius $R_{MT}\approx 2\, r_B$.
The single-site DMFT approximation truncates the self-energy to
$\braket{\Phi^\mu_m | \Sigma |\Phi^{\mu'}_{m'}}=\delta_{\mu,\mu'}\braket{\Phi^\mu_m | \Sigma |\Phi^{\mu}_{m'}}$, and cluster-DMFT keeps
intersite terms $\braket{\Phi^\mu_m | \Sigma |\Phi^{\mu'}_{m'}}$ within the given cluster
(cellular-DMFT~\cite{cellular}). As the functions $\ket{\Phi^\mu_m}$ are
very localized, the negligence of off-site terms on un-dimerized
vanadiums is an excellent approximation.

When dynamic correlations are expanded in the form of Eq.~\ref{Eq:Sigma},
one can not build a low energy (Hubbard-like) model containing only a
small number of bands, because all bands hybridize with atomic degrees
of freedom.  Indeed, the Dyson equation
\begin{eqnarray}
G_{i\omega}(\vr,\vr') = \left((i\omega + \mu +\nabla^{2}-V_{KS}(\mathbf{r}))\delta(\mathbf{r}-\mathbf{r}')
\right.
\nonumber\\
\left.
-\Sigma_{i\omega}(\vr,\vr')\right)^{-1}
\label{Eq:Dyson}
\end{eqnarray}
couples all bands with the atomic-degrees of freedom, and therefore the
Dyson equation has to be solved either in real space, or, equivalently
in complete basis, such as Kohn-Sham basis. This approach is also
called embedded DMFT, as the self-energy is embedded into the large
Hilbert space, rather than the more common truncation of the degrees
of freedom of the solid to a Hubbard model through downfolding.

The truncation to local Feynman diagrams (which is the DMFT
approximation) is less severe
approximation in such formulation of the theory than in the
alternative Hubbard model formulation. Another advantage of such
formulation of DFT+DMFT is that the approximation is conserving and
stationary~\cite{Freeenergy-Haule}, as it is derivable from a
Luttinger-Ward (LW)
functional.
Namely, LW functional
\begin{eqnarray}
\Gamma[G] = \Tr\log(-G) -\Tr((G_0^{-1}-G^{-1})G) + 
\nonumber\\
E_{H+XC}[\rho]+\Phi[G_{local}]-E_{H+XC}[\rho_{local}]
\end{eqnarray}
is stationary at the DFT+DMFT solution, i.e., $\delta
\Gamma[G]/\delta G=0$ at $G=G_{DMFT}$.

To prove that, we take the derivative of the above functional, and obtain
\begin{eqnarray}
\frac{\delta\Gamma}{\delta G}=G^{-1}-G_0^{-1} + V_{KS} + \frac{\delta\tilde{\Phi}[G_{local}]}{\delta G}, 
\end{eqnarray}
where we denoted $\tilde{\Phi}[G_{local}]=\Phi[G_{local}]-E_{H+XC}[\rho_{local}]$.
This derivative vanishes if the self-energy in the Dyson equation
(Eq.~\ref{Eq:Dyson}) satisfies
\begin{eqnarray}
\Sigma = \frac{\delta \tilde{\Phi}[G_{local}]}{\delta G} = 
\frac{\delta G_{local}}{\delta G}\frac{\delta \tilde{\Phi}[G_{local}]}{\delta G_{local}}
\end{eqnarray}
In our embedded formulation of DMFT, $G_{local}$ is a projection of the Green's function to the set
of quasi-atomic orbitals
\begin{eqnarray}
G_{local}(\vr,\vr') = 
\hspace{-0.7cm}
\sum_{\begin{scriptsize}\begin{array}{c}mm' \\\mu,\mu'\in cluster\end{array}\end{scriptsize}} 
\hspace{-0.8cm}
\braket{\vr|\Phi^\mu_m}
\braket{\Phi^\mu_m | G |\Phi^{\mu'}_{m'}}
\braket{\Phi^{\mu'}_{m'}|\vr'}.
\end{eqnarray}
Using this form of the projector, we can compute 
${\delta  G_{local}}/{\delta G}$, and hence the form of the self-energy
at the stationary DFT+DMFT solution:
\begin{eqnarray}
\Sigma(\vr,\vr') = 
\hspace{-0.7cm}
\sum_{\begin{scriptsize}\begin{array}{c}mm' \\\mu,\mu'\in cluster\end{array}\end{scriptsize}} 
\hspace{-0.8cm}\braket{\vr|\Phi^\mu_m}
\braket{\Phi^\mu_m | \frac{\delta \tilde{\Phi}[G_{local}]}{\delta G_{local}}|\Phi^{\mu'}_{m'}}
\braket{\Phi^{\mu'}_{m'}|\vr'}
\label{Eq:Sigma2}
\end{eqnarray}
This is exactly the form postulated in Eq.~\ref{Eq:Sigma}, hence we
proved that DFT+DMFT set of equations are derivable from the $\Gamma$
functional, and are stationary. Note that
within DMFT approximation,
$\frac{\delta \Phi[G_{local}]}{\delta G_{local}}$ is computed by
solving an appropriate quantum impurity model.  Note also that in
deriving Eq.~\ref{Eq:Sigma2} we assumed that the basis functions
$\ket{\Phi^\mu_m}$ do not depend on the Green's function (or the
charge density), and hence $\delta\ket{\Phi^\mu_m}/\delta G=0$.  In
the alternative downfolding implementations of charge self-consistent
DFT+DMFT, the $\ket{\Phi^\mu_m}$ functions are replaced by the Wannier
functions, which essentially depend on the charge density of the
solid, and therefore on the Green's function. In such Wannier
implementations, the derivative of the projector (originating in
nonzero $\delta\ket{\Phi^\mu_m}/\delta G$) has always been neglected,
and therefore charge-self consistent DFT+DMFT method is non-stationary
in Wannier implementations.

\section{Local and Intersite Self energies}

Within our approach the cluster self-energy is diagonal in the
momentum basis, with quantities $\Sigma_{K=0}$ and $\Sigma_{K=\pi}$.
The cluster self-energy in the site representation, on the
other hand, contains the local and intersite self-energies in the following
form \\
\begin{center}
$ 
\Sigma_{c} = 
\begin{pmatrix}
 \Sigma_{11} & \Sigma_{12} \\
\Sigma_{21} & \Sigma_{22}  \\ 
\end{pmatrix}. 
$
\end{center}
Using the cluster symmetry one can notice that $\Sigma_{11} = \Sigma_{22}$ and $\Sigma_{12} = \Sigma_{21}$, where the former is the local self-energy and the later the intersite self-energy. By means of a Fourier transform one can obtain the self-energy in the momentum basis from the local and intersite self-energies~\cite{reviewcluster}, 
\begin{equation}
  \Sigma_{K} = \frac{1}{N_c}\sum_{i,j} \Sigma_{ij} e^{iK(R_{i}-R_{j})},
\end{equation}
where $N_{c}$ is the number of sites within the cluster, $K$ is the cluster momentum, and $R_{i,j}$ denote the sites within the cluster. In our cluster we have $R = 0,1$ and $K = 0,\pi$. Thus, one obtain that 
\begin{equation}
 \Sigma_{K=0} = \Sigma_{11}+\Sigma_{12},
\end{equation}
and
\begin{equation}
 \Sigma_{K=\pi} = \Sigma_{11}-\Sigma_{12}.
\end{equation}
Based on the molecular scenario with bonding and antibonding orbitals, we change the names of $\Sigma_{K=0}$ and $\Sigma_{K=\pi}$ to bonding ($\Sigma_{b}$) and antibonding ($\Sigma_{ab}$) self-energies, respectively. Therefore, the two previous equations lead to the following linear relations:
\begin{equation}
 \Sigma_{11} = \frac{1}{2}(\Sigma_{b} + \Sigma_{ab}),
\label{eq:onsite}
\end{equation}
\begin{equation}
 \Sigma_{12} = \frac{1}{2}(\Sigma_{b} - \Sigma_{ab}),
\label{eq:intersite}
\end{equation}
which are used in the main part of the text.

As we showed in the main text, the imaginary part of the self-energies
associated with the bonding and antibonding states in the M$_1$ phase
vanishes at the Fermi level.  Hence, we can discuss the
bonding-antibonding splitting within an effective band structure
picture, in which the effects of electronic correlations are taken
into account expanding the self-energies to first order. In this case
the bonding and antibonding bands are renormalized by the quantities
\begin{equation}
 Z_{b} = \Bigg( 1- \frac{\partial \Re\Sigma_{b}}{\partial \omega}\Bigg|_{\omega=0}\Bigg)^{-1},
\label{eq:zb}
\end{equation}
and
\begin{equation}
 Z_{ab} = \Bigg( 1- \frac{\partial \Re\Sigma_{ab}}{\partial \omega}\Bigg|_{\omega=0}\Bigg)^{-1}.
\label{eq:zab}
\end{equation}
The value of these quantities are well defined since there is no pole 
at zero frequency in the Matsubara self-energy. As a result,
we can obtain the energies of renormalized bands
\begin{equation}
 E_{b} = Z_{b}(\epsilon_{b}+ \Re\Sigma_{b}(\omega=0)),
\end{equation}
\begin{equation}
 E_{ab} = Z_{ab}(\epsilon_{ab}+\Re\Sigma_{ab}(\omega=0)),
\end{equation}
where $\epsilon_{b}$ and $\epsilon_{ab}$ are the energies of noninteracting bonding and antibonding bands, respectively.

Therefore, the bonding-antibonding splitting energy, $\Delta = E_{ab}-E_{b}$, can be written as 
\begin{align}
\Delta = & Z_{ab}(\epsilon_{ab}+\Re\Sigma_{ab}(0))-Z_{b}(\epsilon_{b}+\Re\Sigma_{b}(0)) \nonumber \\
       = & \frac{(Z_{ab}-Z_{b})}{2}(\epsilon_{ab}+\Re\Sigma_{ab}(0)+\epsilon_{b}+\Re\Sigma_{b}(0)) \nonumber \\+  & \frac{(Z_{ab}+Z_{b})}{2}(\epsilon_{ab}+\Re\Sigma_{ab}(0)-\epsilon_{b}-\Re\Sigma_{b}(0)).
\end{align}

Taking the bonding and antibonding self-energies in terms of $\Sigma_{11}$ and $\Sigma_{12}$ we obtain
\begin{align}
 Z_{b}= \frac{(1-\frac{\partial \Re\Sigma_{11}}{\partial \omega}|_{\omega=0})+(\frac{\partial \Re\Sigma_{12}}{\partial \omega}|_{\omega=0})}{(1-\frac{\partial \Re\Sigma_{11}}{\partial \omega}|_{\omega=0})^{2}-(\frac{\partial \Re\Sigma_{12}}{\partial \omega}|_{\omega=0})^{2}},
\label{eq:Zsum}
\end{align}
\begin{align}
 Z_{ab}= \frac{(1-\frac{\partial \Re\Sigma_{11}}{\partial \omega}|_{\omega=0})-(\frac{\partial \Re\Sigma_{12}}{\partial \omega}|_{\omega=0})}{(1-\frac{\partial \Re\Sigma_{11}}{\partial \omega}|_{\omega=0})^{2}-(\frac{\partial \Re\Sigma_{12}}{\partial \omega}|_{\omega=0})^{2}}.
\label{eq:Zdiff}
\end{align}
Notice that since $Z_{ab}$, $Z_b$, and 
$(1-\frac{\partial \Re\Sigma_{11}}{\partial \omega}|_{\omega=0})$
are positive, it is the sign of $\frac{\partial
  \Re\Sigma_{12}}{\partial \omega}|_{\omega=0}$ that determines which
orbital is more correlated.

Finally, on the basis of equations~(\ref{eq:Zsum}) and~(\ref{eq:Zdiff}) the bonding-antibonding splitting energy is then given by 
\begin{align}
 \Delta = & \frac{(1-\frac{\partial \Re\Sigma_{11}}{\partial \omega}|_{\omega=0})}{(1-\frac{\partial \Re\Sigma_{11}}{\partial \omega}|_{\omega=0})^{2}-(\frac{\partial \Re\Sigma_{12}}{\partial \omega}|_{\omega=0})^{2}} \nonumber \\ & \times (\epsilon_{ab}-\epsilon_{b}-2\Re\Sigma_{12}(0)) \nonumber \\ & - \frac{(\frac{\partial \Re\Sigma_{12}}{\partial \omega}|_{\omega=0})}{(1-\frac{\partial \Re\Sigma_{11}}{\partial \omega}|_{\omega=0})^{2}-(\frac{\partial \Re\Sigma_{12}}{\partial \omega}|_{\omega=0})^{2}} \nonumber \\ & \times (\epsilon_{ab}+\epsilon_{b}+2\Re\Sigma_{11}(0)).
\end{align}
This formula has a simple physical interpretation. The first term is
just the ``local'' $Z_{local}$ multiplied by the bare splitting,
enhanced by the intersite self-energy, while the second term arises
from the asymmetry of the local crystal field levels corrected by
local self-energy, and would vanish in the particle-hole symmetric limit.